\documentclass[conference]{IEEEtran}
\IEEEoverridecommandlockouts
\usepackage[
  pass,
]{geometry}
\usepackage{graphicx}
\usepackage{enumitem}
\usepackage{xcolor,cite,balance}
\usepackage{algorithm}
\usepackage{algpseudocode}
\usepackage{float}
\usepackage{subcaption}
\usepackage{caption}
\usepackage{mdframed}
\usepackage{algorithm}
\usepackage{algpseudocode}
\usepackage{blindtext}
\usepackage[breakable]{tcolorbox}
\usepackage{amsmath}
\usepackage{cleveref}
\crefformat{section}{\S#2#1#3}

\usepackage{authblk}

\title{Towards Scalable Quantum Repeater Networks}

\author{
    Connor Howe\textsuperscript{1}, Mohsin Aziz\textsuperscript{2}, Ali Anwar\textsuperscript{1} \\
    \textsuperscript{1}Department of Computer Science and Engineering, University of Minnesota-Twin Cities\\
    \textsuperscript{2}Ericsson Canada Inc. \\
    Email: \{howe0427, aanwar\}@umn.edu, azizmohsin@live.com
}

\date{\vspace{-5ex}}

\begin{document}
\maketitle

\begin{abstract}
This paper presents a comprehensive study on the scalability challenges and opportunities in quantum communication networks, with the goal of determining parameters that impact networks most as well as the trends that appear when scaling networks. We design simulations of quantum networks comprised of router nodes made up of trapped-ion qubits, separated by quantum repeaters in the form of Bell State Measurement (BSM) nodes. Such networks hold the promise of securely sharing quantum information and enabling high-power distributed quantum computing. Despite the promises, quantum networks encounter scalability issues due to noise and operational errors. Through a modular approach, our research aims to surmount these challenges, focusing on effects from scaling node counts and separation distances while monitoring low-quality communication arising from decoherence effects. We aim to pinpoint the critical features within networks essential for advancing scalable, large-scale quantum computing systems. Our findings underscore the impact of several network parameters on scalability, highlighting a critical insight into the trade-offs between the number of repeaters and the quality of entanglement generated. This paper lays the groundwork for future explorations into optimized quantum network designs and protocols.
    
\end{abstract}

\section{Introduction}

Quantum communication networks distinguish themselves from classical networks by leveraging the fundamental principles of quantum mechanics, offering a fundamentally different and potentially superior form of information exchange. Despite their promising advantages, quantum communication networks face significant scalability challenges \cite{PhysRevResearch.3.013279}. These challenges arise from various issues, including decoherence, which leads to the loss of quantum information over time, and other factors such as photon loss (specifically for trapped-ion and photonic systems) during transmission, and the limited entanglement times \cite{Chen2023Scalable}.

One promising approach to addressing these challenges is the development of quantum repeaters. Quantum repeaters are analogous to classical repeaters in that they serve to extend communication distances, but they operate on completely different principles \cite{Briegel1998Quantum}. Instead of simply amplifying signals, quantum repeaters create intermediate nodes that perform entanglement swapping and purification techniques to link distant qubits. This process effectively mitigates photon loss and decoherence, allowing for longer-distance quantum communication \cite{PhysRevLett.128.150502}.

In a successful implementation of a quantum network \cite{mit_harvard_2020}, researchers at MIT and Harvard developed a quantum repeater using silicon-vacancy centers in diamond as quantum memory modules. These repeaters demonstrated the ability to maintain entanglement over 50 kilometers of optical fiber, showcasing the potential for scalable quantum networks. These repeaters work by entangling photons emitted from qubits and performing entanglement swapping to extend the entanglement range across multiple nodes.

A team at the University of Innsbruck successfully demonstrated entanglement between routers comprised of trapped ion qubits. This entanglement was also done over 50 kilometers using calcium ions entangled with photons, which were then transmitted through optical fibers to distant nodes. This setup achieved high fidelity and a significant entanglement success rate, highlighting the feasibility of trapped ions in practical quantum repeater systems \cite{innsbruck_2023}.

In the domain of quantum communication, there is a significant enhancement in the capacity for distributed storage and processing of information. This is achieved by facilitating an efficient and secure exchange of quantum information, leading to the possibility of using quantum processing technologies for increased computational power and the development of more fault-tolerant systems. The current generation of quantum computing hardware, referred to as Noisy Intermediate-Scale Quantum (NISQ) devices, is operational but faces limitations due to noise and errors during quantum operations~\cite{bharti2022noisy, stilck2021limitations}. The implementation of advanced quantum communication techniques to scale up NISQ devices is pivotal for enabling more complex quantum circuits and fostering communication between quantum processors, thereby unlocking new capabilities and applications in quantum computing~\cite{wehner2018quantum}.

However, a significant challenge remains in the scalability of quantum communication networks. Existing solutions often face limitations in their ability to scale efficiently~\cite{caleffi2022distributed}. This paper aims to address these challenges by adopting a modular approach towards the development of scalable, large-scale quantum computing systems. Our focus is on analyzing important trends and parameters for scaling quantum networks and identifying areas of focus for the practical implementation of such networks.

The contributions of this paper are as follows:
\begin{itemize}
    \item{We explore the limits to scalability in quantum communication networks by analyzing the entanglement generation rate and end-to-end qubit fidelity.}
    \item{Building upon the SeQUeNCe framework, we develop new simulations of adaptable trapped-ion qubit quantum communication networks.}
    \item{Using our simulated networks, we show trends in entanglement generation and end-to-end fidelities in order to identify important considerations when scaling quantum networks.}
\end{itemize}

\section{Related Work}
The transition from classical to quantum network evaluation methodologies marks a pivotal evolution in the field of communication networks. Initially, research in classical networks established a foundation by focusing on critical performance metrics such as throughput, latency, and packet loss. Techniques developed for optimizing throughput in diverse wireless network systems utilized a broad spectrum of algorithms\cite{Lashgari2013TimelyThroughput}. Moreover, in the realm of latency optimization, significant contributions have detailed the use of Software Defined Networking (SDN) for dynamically routing internet traffic along paths with the lowest latency, thus enhancing the reliability of time-sensitive applications\cite{Llopis2016MinimizingLatency}. Drawing from the foundational work of Lashgari on throughput optimization in heterogeneous wireless networks and Llopis on latency minimization in SDN contexts, we aim to extend these core concepts—specifically real-time routing optimization and dynamic traffic management—to the quantum domain. This entails exploring their applicability and the necessary adjustments to meet the unique challenges and leverage the distinct capabilities of quantum networks. Our current paper will open up avenues to focus on this translation in future works.

Optimizing network performance to minimize packet loss remains a vital area of research, offering valuable insights for the development of optimization algorithms\cite{GhaniAlJobouri2023PacketLoss,LanjewarGuptaAODV}. Although such algorithms may not directly transfer to the quantum setting due to the fundamentally different nature of quantum phenomena, they lay a conceptual groundwork for devising strategies to minimize decoherence. These strategies, crucial for the forward march of quantum networking technologies, are however beyond the purview of this paper.

Classical network simulations have been instrumental in assessing the reliability and deployability of internet protocols\cite{BreslauEstrinFallFloydHeidemannHelmyHuangMcCanneVaradhanXuYu2023Advances}, playing a significant role in the early stages of network development. In the quantum domain, the SeQUeNCe framework marks a significant step forward in developing quantum simulation tools for quantum network environments, underscoring the challenges and necessities of simulating quantum-specific phenomena\cite{wu2021sequence}. This foray into quantum simulations necessitates a reassessment of classical methodologies, many of which do not directly apply. We leverage the SeQUeNCe framework\cite{wu2021sequence} for simulating quantum networks, utilizing its extensive capabilities to model and analyze the complex behaviors and protocols unique to quantum communication systems accurately.

Historically, quantum network research has concentrated on designing robust physical hardware for quantum networks\cite{pant2017rate, azuma2022quantum}. Quantum memories, for instance, have demonstrated potential in facilitating quantum information transfer across larger, more complex networks, mitigating the need for uniform channel delay\cite{Wang2022QuantumMemory}. Additionally, there has been a focused effort on mitigating the effects of decoherence and errors in quantum communication processes, with various qubit technologies being tested and analyzed to identify the most effective methods for combating decoherence. Recent advancements in quantum network systems have explored the implementation of trapped-ion qubits\cite{Drmota2023QuantumMemory}, photonic qubits\cite{Bogdanov2019QuantumPlasmonics}, and the theoretically promising topological qubits\cite{Wille2019Majorana}. These hardware designs and system implementations often stem from the challenges of quantum network scalability, prompting an exploration of quantum information's inherent properties and its environmental interactions. This exploration has led to research addressing scalability limitations through routing optimization\cite{chehimi2023matching}, new entanglement swapping protocols\cite{huang2022entanglement}, and the optimization of distillation resources\cite{rozpkedek2018optimizing}, significantly enhancing quantum networks' performance and scalability. However, the study of scalability limits in conjunction with optimized routing and diverse hardware remains underexplored. Current work investigates how entanglement generation rate and qubit fidelity vary in linear, homogeneous quantum networks as the total distance spanned by the network increases\cite{chehimi2023scaling}. We aim to build on this research by extending our analysis to non-linear, heterogeneous networks, incorporating a wide array of hardware specifications, and exploring multiple network configurations.

The paper "Scaling Limits of Quantum Repeater Networks" by Mehdi Cheimi \cite{chehimi2023scaling, chehimi2023matching} explores the challenges in scaling quantum repeater networks, which are essential for long-distance quantum communication. The study focuses on the trade-offs between total network distance and the overall performance of the network. Cheimi introduces a framework to analyze the capacity of such networks, considering factors like error rates and the time needed for entanglement generation. The findings indicate that while increasing the number of repeaters can improve network performance, there are diminishing returns due to the complexity and resource requirements, highlighting the need for optimized network design to achieve efficient and scalable quantum communication. The paper also does not account for heterogeneous networks, where variables such as different distance separations, varying topologies, and differing repeater counts might exist across the network. In another work, they did address heterogeneous networks for entanglement rate optimization, but this paper aims to expand on these results and identify more scaling trends \cite{Chehimi2021Entanglement}.

\section{Methodology}

In this section, we outline the methodology employed to establish and monitor simulated quantum networks, and we describe the experimental design aimed at identifying the scalability limits of these networks.

\subsection{Network Setup}

Our quantum network simulation is designed to represent an advanced quantum communication system, incorporating quantum routers and BSM nodes interconnected through both classical and quantum optical channels. These nodes work together to facilitate complex processes involved in establishing and manipulating quantum states across a distributed network.

Quantum routers, equipped with 50 trapped ion qubits each, act as the primary nodes that initiate and control quantum operations such as entanglement generation and swapping. In the entanglement generation process, a photon is emitted from a qubit in each of the two quantum routers. These photons travel to the intermediate BSM node, where they are subject to a joint measurement. The single-atom BSM node detects the incoming photons and performs a Bell State Measurement. This process projects the qubits in the quantum routers into an entangled state, effectively creating a link between the routers. By repeating this procedure across the network, we can extend the range and connectivity of the quantum network.

At the single-atom BSM nodes, entanglement swaps can be performed. In this process, photons from two previously entangled pairs are jointly measured at the BSM node, effectively entangling the two remote qubits that were not directly interacting. This allows for the extension of entanglement across multiple nodes, significantly increasing the reach of the quantum network. 

\begin{figure}[h]
    \centering
    \includegraphics[width=0.9\linewidth]{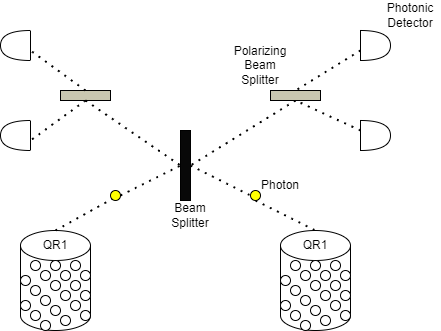}
    \caption{Entanglement generation process between two quantum routers, facilitated at the intermediary BSM node. Shows basic network setup and flow of network processes.}
    \label{Figure 1}
\end{figure}

We designate two distinct communication channels within our quantum network architecture: one for classical communication and one for quantum communication. The classical communication channel is responsible for transmitting information necessary for scheduling and requesting entanglements. This includes coordination signals, entanglement requests, and acknowledgment messages, ensuring that the network operates efficiently and synchronously.

The quantum communication channel, on the other hand, is dedicated to the transfer of photons through fiber optic cables. This channel handles the delicate process of quantum information transfer, where photons emitted from qubits are transmitted to facilitate entanglement between distant nodes. The use of fiber optic cables minimizes photon loss and maintains the integrity of quantum states over long distances. This dual-channel approach ensures that the classical and quantum information flows are optimized for their respective roles, enhancing the overall performance and reliability of the quantum network.

\begin{figure}[h]
    \centering
    \includegraphics[width=0.95\linewidth]{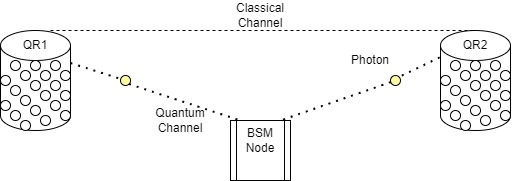}
    \caption{General network setup, showing entanglement between two routers.}
    \label{Figure 1}
\end{figure}

The simulation employs the Barrett-Kok entanglement generation protocol. This protocol manages key parameters such as quantum delay, frequency of local memory excitation, emission times, measurement results, success probability, and fidelity degradation, all of which are critical for understanding network scalability.

Dynamic management of the network includes protocols for state initialization, resource allocation, and real-time adjustments based on event responses. Quantum states are initialized and updated across nodes based on generation protocol requests and results from quantum measurements and channel transmissions. The network also allocates and manages photons dynamically to maintain high fidelity and efficiency in operations.

The simulation is highly customizable, allowing users to modify operational parameters such as transmission frequencies, channel attenuation, and memory coherence times. This adjustability, combined with the capability to scale the network in terms of node count, channel length, and protocol complexity, makes the simulation a robust platform for exploring the potentials and limitations of quantum repeater networks. This framework serves not only as a tool for understanding fundamental quantum communication principles but also as a testing ground for advanced quantum protocols and network configurations under controlled, simulated conditions.

\subsection{Analysis Methodology}

To determine scalability in quantum repeater networks, we adjusted parameters in a controlled environment, aiming to extract metrics like end-to-end entanglement and fidelity across communication nodes. Our Quantum Network Simulation Algorithm, as detailed in Algorithm \ref{alg:QuantumNetworkSimulation}, is crucial for this purpose. It models interactions between quantum nodes using inputs such as node count, channel characteristics, and qubit coherence times. The simulation manages tasks like initializing routers, establishing communication channels, and executing quantum operations such as entanglement swapping. The results, including entanglement count and fidelity, offer valuable insights into network performance and the effectiveness of quantum communication protocols. This approach enhances our understanding and optimization of quantum networks.



\begin{algorithm}[H]
\caption{Quantum Network Simulation Algorithm for Distributed Quantum Systems}
\begin{small}
\begin{algorithmic}[1] 
    \State \textbf{Input:} Simulation time $T_{sim}$, channel delay $\delta_{channel}$, channel attenuation $\alpha_{channel}$, node separation distances $d_{nodes}$, number of nodes $N$, Qubit coherence times $\tau_{coh}$
    \State \textbf{Output:} Entanglement count $E_{count}$, end-to-end fidelity $F_{e2e}$

    \Statex \textit{Initialize simulation parameters}
    \State $T \gets$ Simulation timeline with duration $T_{sim}$
    \State Discretize time with granularity $\Delta t$
    \State Initialize event queue for quantum operations

    \Function{Network\_Setup}{$N$, $\delta_{channel}$, $\alpha_{channel}$, $d_{nodes}$, $\tau_{coh}$}
        \State \textbf{Initialize $N$ quantum routers $\{R_i\}_{i=1}^N$}
        \For{each router $R_i$ in the network}
            \State Assign unique ID to router $R_i$
            \State Initialize quantum memory $M_i$ with coherence time $\tau_{coh}$ and initial fidelity $F_{init}$
        \EndFor

        \State \textbf{Create $N-1$ Bell State Measurement (BSM) nodes}
        \For{each adjacent router pair $(R_i, R_{i+1})$}
            \State Connect router $R_i$ and $R_{i+1}$ via a BSM node
        \EndFor

        \State \textbf{Establish classical communication channels}
        \For{each classical channel $C_c(i,j)$}
            \State Configure delay $\delta_{channel}(i,j)$
        \EndFor

        \State \textbf{Establish quantum communication channels}
        \For{each quantum channel $C_q(i,j)$}
            \State Configure channel attenuation $\alpha_{channel}(i,j)$ and set node separation $d_{nodes}(i,j)$
        \EndFor

        \State \textbf{Update quantum memory parameters in each router}
        \For{each router $R_i$}
            \State Model coherence decay and fidelity degradation over time
        \EndFor
    \EndFunction

    \State \textbf{Configure entanglement swapping rules}
    \State Define protocols for entanglement swapping and routing
    \State Define routing table

    \State \textbf{Initialize quantum network management entity $M_n$}
    \State Manage entanglement requests, routing decisions, and resource allocation

    \State \textbf{Start simulation timeline $T$}
    \State Execute quantum operations, including entanglement generation and swapping

    \State \textbf{Process simulation results}
    \State Monitor and log outcomes such as entanglement generation success
    \State Record fidelity degradation and successful swapping events

    \State \textbf{Output simulation results}
    \State Compute total entanglement count $E_{count}$ and end-to-end fidelity $F_{e2e}$
    \State Store results for analysis and visualization

\end{algorithmic}
\end{small}
\label{alg:QuantumNetworkSimulation}
\end{algorithm}

In our quantum network setup, we employ an adaptation of the Barrett-Kok entanglement generation protocol, to facilitate remote entanglement between qubits stored in quantum routers. This protocol operates as follows \cite{PhysRevA.71.060310}:

\begin{enumerate}
    \item \textbf{Photon Emission and State Preparation:} Each quantum memory is prepared in a superposition state \( |+ \rangle \) and excited to emit a photon that remains entangled with the qubit. We ensure controlled photon emission towards an intermediary BSM node.
    
    \item \textbf{Bell State Measurement:} Photons transmitted from separate routers converge at the BSM node, where their entanglement is measured. The result of this BSM determines the entanglement state of the distant qubits and is crucial for the subsequent steps in the protocol.
    
    \item \textbf{Feedback and Correction:} Based on the BSM outcomes, classical communications are sent back to the originating routers to apply necessary quantum corrections. Depending on the measurement results, Pauli X or Z gates are selectively applied to the qubits to correct their states and finalize the entanglement.
\end{enumerate}

This setup forms the core of our experimental entanglement generation, leveraging both quantum and classical channels to achieve robust and verifiable quantum entanglement across the network. The protocol’s efficacy is determined by its ability to synchronize these operations across network components and handle the probabilistic nature of quantum measurements and state corrections. We limit the noise and work with idealized network conditions in order to determine the effects of certain specific parameters such as separation distance and node count. 

\section{Experimental Setup}

Accurately determining the scalability limits of quantum networks requires careful tuning of network parameters. We point to two key metrics for evaluating the simulated networks: entanglement generation rate and end-to-end fidelity. 
We aim to understand the behavior of a quantum network when entanglement generation is requested between two end routers arranged in a linear chain. We label the initial router r0 and the final router rX where X is one less than the total number of routers, defined in the algorithm \ref{alg:QuantumNetworkSimulation}. The experiment is designed to evaluate how different factors, such as total distance and the number of intermediate nodes, influence the success of entanglement generation. Specifically, we investigate the conditions under which the network scales, the impact of various network configurations, and how these factors collectively affect the overall entanglement generation efficiency.

In order to determine the scalability limits of these networks, we start with two changeable variables. 

\begin{itemize}
    \item \textbf{Total Distance}: We vary the total distance between the end routers, considering three scenarios: 100 km, 1,000 km, and 10,000 km. This range allows us to observe the effects of distance on entanglement fidelity and the success rate of swapping operations.
    \item \textbf{Node Count}: For each total distance, we analyze the network's performance by varying the number of intermediate nodes up to a maximum of 17. This allows us to explore how the number of nodes, and hence the number of required swaps, influences the overall entanglement generation.
\end{itemize}

Both distance and node count have significant impacts on the total number of entanglements generated. At greater distances, quantum memories degrade over time, reducing the fidelity of entangled states and making successful swaps less likely. We work with close-to-ideal quantum states, setting decoherence parameters lower than they would be in practical quantum networks in order to observe the networks' scalability limits without the presence of complicated noise. We update coherence times with for quantum memories in the algorithm \ref{alg:QuantumNetworkSimulation}. Additionally, as the number of nodes increases, more swaps are required, each introducing a potential point of failure, especially if the memory fidelities are already low due to long distances between nodes.

In order to determine optimal individual router separations, we start by aligning a certain number of routers in a chain of fixed total distance. In the algorithm \ref{alg:QuantumNetworkSimulation}, we adjust the number of routers in the network of that fixed distance. We begin by examining three vastly different network distances; 100km, 1000km, and 10000km. Within each of these distances, we identify the number of successful entanglements generated between the first and last routers in the chain out of twenty attempts. By exploring scalability this way, we find how networks behave in the extremes (long and short distance spans), and which parameters matter most for determining scalability trends at the extremes. Although the extremes that we observe here are far above that of practical quantum networks, they give us insights into the trends and important parameters in such networks. With this information, we will be able to more accurately develop routing algorithms and schedulers for quantum networks in the future. 

\section{Evaluation}

To guide our analysis and gain deeper insights into the scalability of quantum communication networks, our evaluation is structured around a set of key research questions. These questions are designed to explore the critical factors influencing network performance and scalability. 

\begin{tcolorbox}[breakable,width=0.48\textwidth,title={Research Questions:},boxrule=.3mm,colback=white,coltitle=white,left=.5mm, right=.5mm, top=.5mm, bottom=.5mm]
\begin{itemize}
    \item \emph{How does a homogeneous network scale (\cref{subsec:eval1})}?
    
    \item \emph{How does a heterogeneous fixed distance network perform with different total router counts 
    (\cref{subsec:eval2})}?
    
    \item \emph{Do the networks show the same trends with different total distances (\cref{subsec:eval3})}?
    
    \item \emph{What is the minimum number of routers necessary to generate entanglement (\cref{subsec:eval4})}?
\end{itemize}
\end{tcolorbox}

\subsection{Homogeneous Network Scaling}
\label{subsec:eval1}

We began our investigation by assessing the scalability of simple homogeneous quantum networks, focusing on the total distance spanned while maintaining various levels of fidelity and entanglement generation rates. Our initial findings revealed a predictable, linear relationship, indicating that networks requiring minimum fidelity and exhibiting the lowest entanglement generation rates could span the greatest distances. This outcome was anticipated and is illustrated in Figure 1. 

\begin{figure}[h]
    \centering
    \includegraphics[width=7cm]{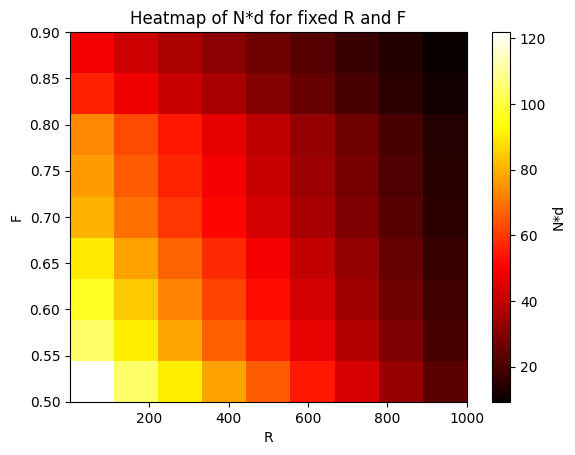}
    \caption{Homogeneous network scalability plot. N is the number of nodes, d distance between nodes, F is end-to-end fidelity and R is the entanglement generation rate.}
    \label{Figure 1}
\end{figure}

Our findings were simple yet practical. However, we wanted to explore some larger extremes and wanted to explore the true scalability limits of quantum repeater networks given controlled scenarios so that we could identify trends and limiting factors in more complex/complete networks. To enhance the complexity and accuracy of our analysis, we subsequently employed the SeQUeNCe framework for more sophisticated simulations, incorporating a broader array of parameters. Through this approach, we simulated heterogeneous networks characterized by non-uniform distance separations and variable node counts, evaluating them based on three critical metrics: end-to-end qubit fidelity, the ratio of failed to successful entanglement attempts, and the total number of successful entanglements versus attempts.

\begin{tcolorbox}[breakable,width=0.48\textwidth,title={Takeaway:},boxrule=.3mm,colback=white,coltitle=white,left=.5mm, right=.5mm, top=.5mm, bottom=.5mm]    
   \emph{Homogeneous networks scale linearly, higher distance corresponds to lower entanglement generation rate and lower end-to-end fidelity.}
\end{tcolorbox}  


\subsection{Heterogeneous Fixed Distance Network}
\label{subsec:eval2}

In our study, we examined the impact of increasing the number of routers and consequently, BSM nodes within a quantum network of fixed spatial distance of 1000km. Our analysis revealed a notable trend: the introduction of additional nodes led to a decrease in the rate of successful quantum entanglements. This observation aligns with theoretical expectations, as each entanglement swapping operation introduces a probability of failure. The results of our experiment for a 1000 km example are displayed in Figure 4. These networks are highly idealized and focus on observing trends in actual network capabilities outside of the presence of high noise, decoherence, or state degradation.  

\begin{figure}[h]
    \centering
    \includegraphics[width=8.75cm]{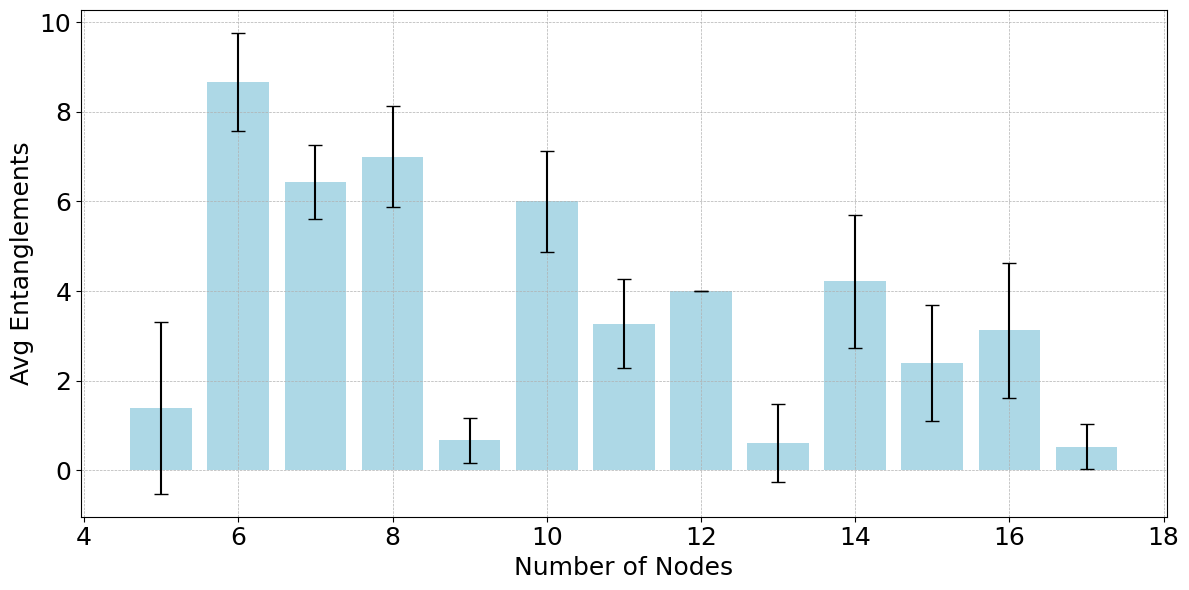}
    \caption{Entanglements generated end-to-end across a simulated quantum network of fixed total distance 1000km while varying node count. We observe three individual trends within the scaling.}
    \label{Figure 4}
\end{figure}

The observed trend emerges when the total distance between nodes falls below a certain threshold, where the risks of failure during entanglement generation outweigh the effects of decoherence. This threshold is a critical metric in quantum networks because it marks an inflection point that leads to fundamentally different patterns of entanglement generation, as demonstrated in subsequent experiments. Understanding this behavior is crucial for the practical implementation of quantum networks, yet it remains under-explored in current discussions. 

\begin{tcolorbox}[breakable,width=0.48\textwidth,title={Takeaway:},boxrule=.3mm,colback=white,coltitle=white,left=.5mm, right=.5mm, top=.5mm, bottom=.5mm]    
   \emph{Complex heterogeneous networks introduce many additional patterns and trends, even without the presence of high noise/decoherence.}
\end{tcolorbox}  

\subsubsection{Even-Odd Scaling}

We notice a secondary trend; that the network displays distinct patterns depending on whether the number of nodes in the chain is even or odd:
\begin{itemize}
    \item \textbf{Even Node Counts}: When the number of nodes is even, swapping occurs at a central BSM node, which provides a symmetric point for performing entanglement swaps, leading to higher success rates. This is also because of the generation protocol used.
    \item \textbf{Odd Node Counts}: With an odd number of nodes, there is no central BSM node, which may disrupt the symmetry and lead to a more complex swapping process under an entirely different generation protocol. This results in decreased entanglement generation efficiency as the network configuration becomes less optimal for centralized swapping.
\end{itemize}

There are three distinct trends within the generation attempts, two within odd node counts and one between odd and even counts. The distinction between even and odd node count generation patterns arises from the network setup itself as well as the nature of how entanglement generation happens. By organizing the nodes in a chain pattern, we subject the network to conditions where requesting entanglement between nodes separated by an even number of routers allows the sent qubits to meet in the middle at a BSM node whereas if they are separated by an odd number of routers they meet at a router where generation is performed differently. 

In scenarios with odd node counts, we see a decrease in entanglement generation as the node count increases. This is particularly true when the quantum memories' fidelities are already low due to large inter-node distances. The greater the number of swaps required, the higher the probability of encountering failures, thus reducing the overall success rate of generating entanglements. This pattern still holds in cases where distances between nodes are sufficiently small, and the fidelities are sufficiently high.

Displayed separately, the patterns within odd node counts appear as follows: 

\begin{figure}[H]
    \centering
    \begin{subfigure}[b]{0.32\textwidth}
        \centering
        \includegraphics[width=\textwidth]{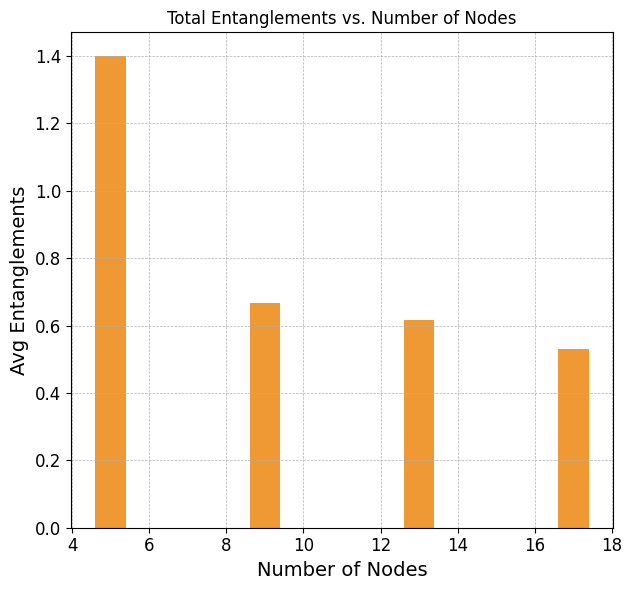}
        \caption{First trend within the odd node counts. Very low average entanglements per node, yet still shows a downward overall trend.}
        \label{fig:odds1}
    \end{subfigure}
    \hspace{0.05\textwidth} 
    \begin{subfigure}[b]{0.32\textwidth}
        \centering
        \includegraphics[width=\textwidth]{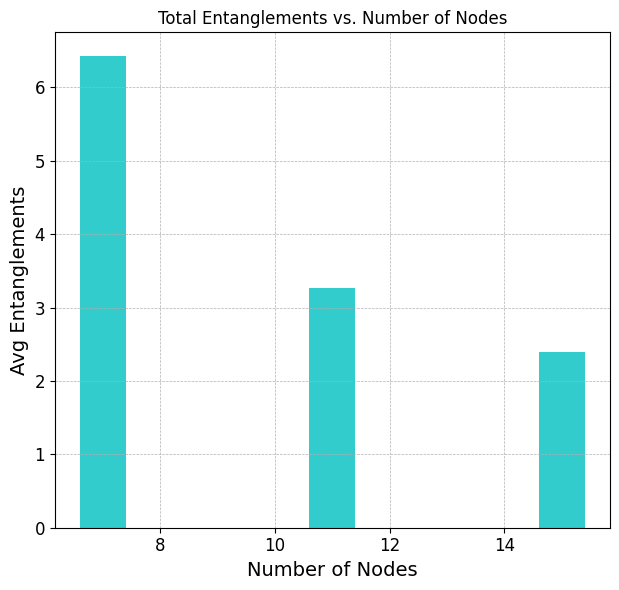}
        \caption{Secondary trend within odd node counts. Entanglements are more successful and a downward trend is still observed.}
        \label{fig:odds2}
    \end{subfigure}
    \caption{Comparison of two separate trends observed within the odd node counts. Within each trend, we still observe the general overall trend yet both vary greatly in a number of entanglements generated.}
    \label{fig:odds-comparison}
\end{figure}

Generation requested between nodes with even router count separation is significantly higher because, at the middle BSM node, the arrival of qubits is synchronized. We also observe a secondary trend within the odd node count runs. This secondary pattern is a result of generation protocols alternating in which BSM node swapping is performed. This results in differing states that the swapping protocols start in, and thus different rates of entanglement generation.

\begin{tcolorbox}[breakable,width=0.48\textwidth,title={Takeaway:},boxrule=.3mm,colback=white,coltitle=white,left=.5mm, right=.5mm, top=.5mm, bottom=.5mm]    
   \emph{Quantum networks use extremely sensitive generation and swapping protocols. Ensuring that those are able to work for each possible configuration is pivotal in maintaining high rates of communication.}
\end{tcolorbox}  

\subsection{Trends across different total distances}
\label{subsec:eval3}

Recognizing these important trends, we extended our analysis to networks with greater total separation distances. Specifically, we examined entanglement generation in networks with total distances of 100 km, 1,000 km, and 10,000 km. We aim to identify whether distance-induced decoherence or swap failure probability contributes most (or at all) to the generation rates at each of these distances. Understanding these bottlenecks is crucial for determining whether hardware limitations (or potentially network routing protocols) are the primary constraints, or if similar trends persist across all distances. We also aim to demonstrate the even-odd scaling trends across these distances in order to highlight the continuous impact that desynchronized entanglements can have.

If we observe a consistent decrease in entanglement generation with increasing node count across all distances, this would suggest that decoherence plays a minimal role in network scalability. However, we know that this is unlikely, even in our highly idealized networks, where we expect decoherence to have a significant impact, particularly at greater distances. On the other hand, if the trend does not persist, and we observe a different pattern, it would indicate that an inflection point has been reached, or that the distances are too short to observe major impacts from decoherence-induced state degradation.   

For a network spanning 100 km, we observe that the distance is sufficiently short to avoid significant decoherence effects. The fidelities of the quantum states remain high, allowing for the expansion to multiple nodes without introducing a substantial increase in failure rates. High fidelity is maintained because the short separation distances between nodes prevent decoherence from degrading the states below the threshold needed for successful operations. As a result, failures in this network primarily arise from random chance or photon loss, rather than from decoherence.

\begin{figure}[H]
    \centering
    \includegraphics[width=8.75cm]{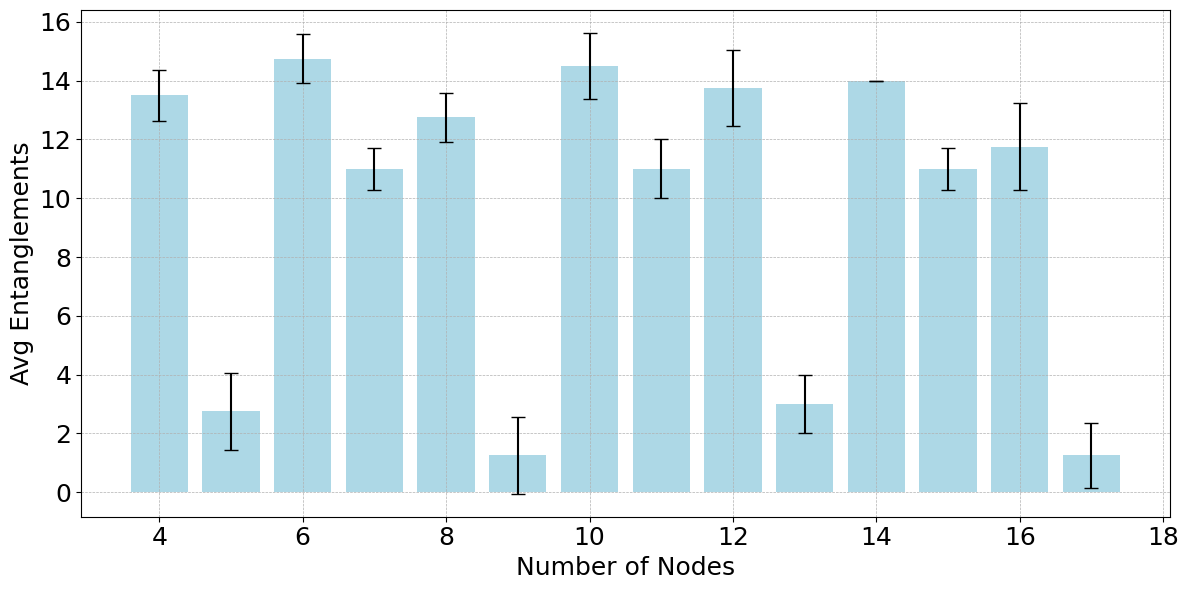}
    \caption{Entanglements generated end-to-end across a 100 km simulated quantum network. We observe no global decrease or increase as we expand node counts, implying that the network is quite stable for the separation distances tested.}
    \label{Figure 4}
\end{figure}

We observe a different trend for our 10000km network. This trend is a mirror of that which we observe for the 1000 km total network distance. This indicates that the network is past the point of decoherence and protocol failure balance whereas the 1000 km example was below that point. Decoherence and state degradation play the largest impact in entanglement failure through the 10000km network and each additional repeater added between the end nodes is able to raise the chance of successful end-to-end entanglement. 

\begin{figure}[H]
    \centering
    \includegraphics[width=8.75cm]{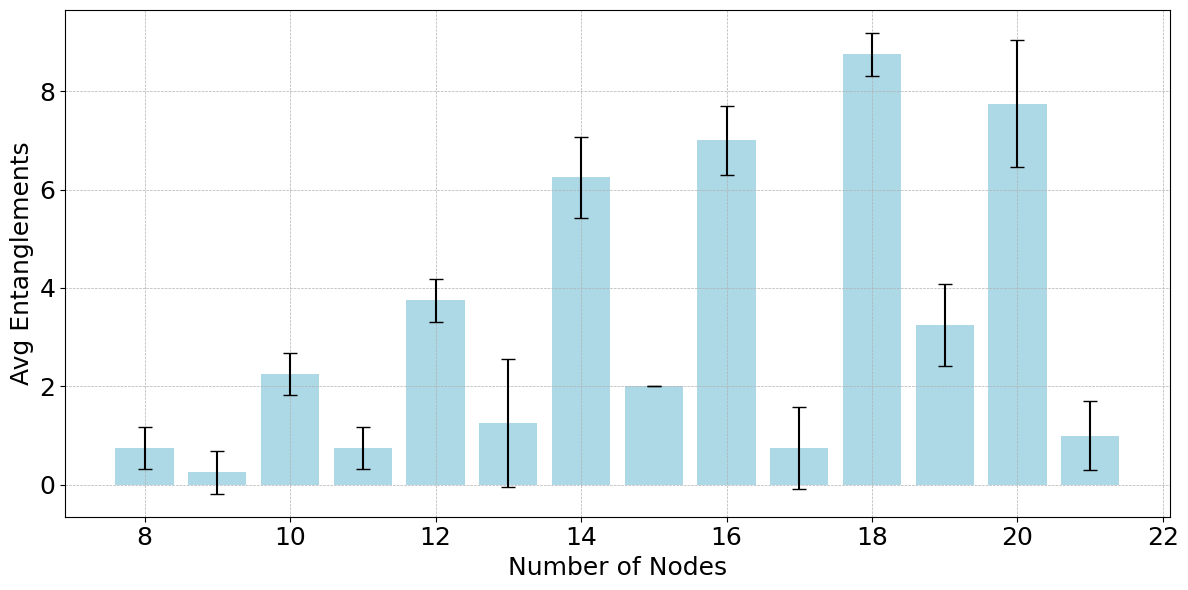}
    \caption{Entanglements generated end-to-end across a simulated quantum network of fixed total distance of 10000km while varying node count. We observe an increase in generation as we add nodes (thus decreasing node separation).}
    \label{Figure 4}
\end{figure}

The observed differences in trends across the networks are attributed to the changing dominant source of failure at different distances. In the 100 km network, the total distance is too short for decoherence to significantly impact the photons as they travel to intermediary BSM nodes. As a result, the swapping and generation procedures begin with high initial fidelities, leading to a low probability of failure, regardless of the number of swaps involved.

In contrast, for the 1,000 km network, the greater distance between nodes causes the quantum states to degrade to a point where successful swaps and entanglement generation are less likely. However, this degradation is not severe enough for decoherence alone to cause entanglement failures. Instead, the generation protocol fails more frequently due to the suboptimal states resulting in more protocol retries and subsequently more failures. 

For the 10,000 km network, decoherence becomes the primary limiting factor. Photons are more likely to not reach their adjacent repeaters than they are to fail in entanglement generation if they do reach them. In this case, adding more routers to reduce the distance between adjacent nodes improves the likelihood of successful entanglement attempts, as the shorter distances help mitigate the impact of decoherence, even if the failure rates at each BSM node remain a concern.

\begin{tcolorbox}[breakable,width=0.48\textwidth,title={Takeaway:},boxrule=.3mm,colback=white,coltitle=white,left=.5mm, right=.5mm, top=.5mm, bottom=.5mm]    
   \emph{Generating entanglements over a large network (or one that experiences high levels of decoherence) requires more repeaters, but the introduction of too many repeaters within a specified distance can be harmful to overall generation.}
\end{tcolorbox}  

\subsection{Minimum repeater counts for large distance networks}
\label{subsec:eval4}

We also analyze the minimum repeater counts under which we observe any entanglement across a network of a set distance. We observe that there is a steady increase as the total distance goes up. This confirms that as decoherence becomes more of a problem, the network requires more repeaters to purify degraded states and extend communication. 

\begin{figure}[H]
    \centering
    \includegraphics[width=8cm]{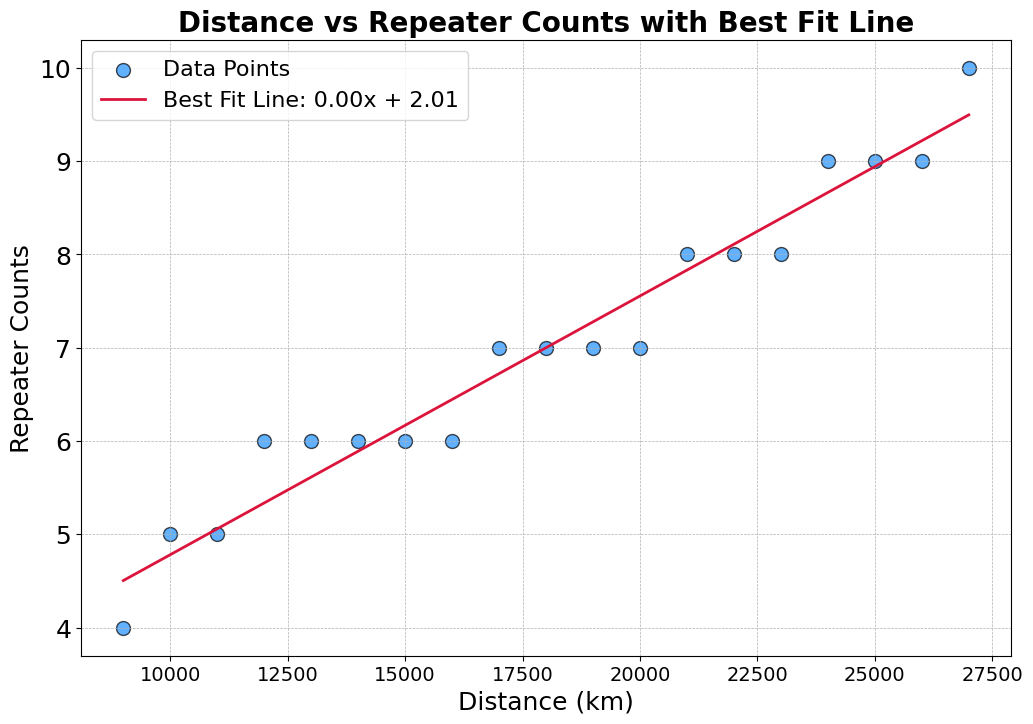}
    \caption{Simple linear fit to scaling trend of minimum repeater count needed to observe at least one entanglement at certain distances.}
    \label{Figure 4}
\end{figure}

The relationship between minimum necessary repeaters and total distance can be approximated by a linear relationship, where the slope is the average minimum distance separation between adjacent routers. This is a valuable metric in determining the maximum range for which routers can be separated while still maintaining communication with the rest of the network.
We use linear regression to find a rough approximation of 3600 km the maximum distance separation per node in our idealized network. Similar techniques could be employed in physical quantum networks to determine what the maximum distance separation can be and to ensure that all routers are kept reachable within the networks.

\begin{tcolorbox}[breakable,width=0.48\textwidth,title={Takeaway:},boxrule=.3mm,colback=white,coltitle=white,left=.5mm, right=.5mm, top=.5mm, bottom=.5mm]    
   \emph{A useful metric for planning practical networks will be maximum node separation (maximum distance that you can keep between two quantum repeaters while still generating entanglement).}
\end{tcolorbox}  

While analyzing the minimum repeaters required for entanglement generation across different network configurations provides a general overview and some important insights into maximum separation, it does not fully capture the complexities and variations inherent in each success. To gain a deeper understanding of the network's performance, it is crucial to examine the average end-to-end fidelity after each successful entanglement attempt. This approach allows us to assess the fidelity of individual links and repeaters, revealing the specific impact of each network component on the overall quality of entanglement. By focusing on the fidelity metrics of these individual entanglement attempts, we can more accurately identify the factors that influence the robustness and reliability of quantum communications within various network configurations.

\begin{figure}[H]
    \centering
    \includegraphics[width=8cm]{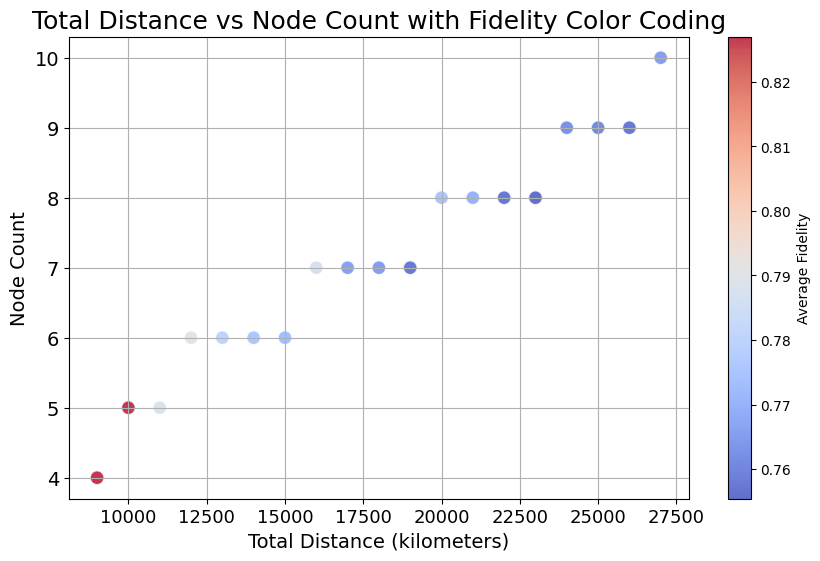}
    \caption{End-to-end fidelities for different repeater counts. Displayed for different minimum repeater counts needed for entanglement generation. We step up by one node once fidelity is too low for entanglement.}
    \label{Figure 4}
\end{figure}

In analyzing the relationship between total distance and node count with respect to end-to-end fidelities, we observe a significant trend where average fidelity tends to decrease as the total distance increases. This decline is punctuated by periodic recoveries in fidelity, particularly noticeable at points where the fidelity approaches a critical threshold of around 0.75. Each time fidelity drops to this critical level, there is a slight increase, which corresponds to the jump in total repeater count. However, these recoveries are followed by a continued downward trend, where the fidelity gradually decreases again, illustrating a progressive challenge in maintaining high fidelity over increasing distances. This pattern underscores the limitations of current quantum entanglement technologies over extended distances, highlighting the need for advanced hardware or network design solutions to enhance fidelity in quantum networks. Notably, each successive recovery fails to restore fidelity to levels observed in previous cycles, indicating a diminishing return despite increased efforts to counteract fidelity loss. This observation is critical for planning and optimizing the infrastructure of quantum communication networks.

\begin{tcolorbox}[breakable,width=0.48\textwidth,title={Takeaway:},boxrule=.3mm,colback=white,coltitle=white,left=.5mm, right=.5mm, top=.5mm, bottom=.5mm]    
   \emph{Fidelity consistently decreases within a network, and by adding repeaters when fidelity becomes low enough that entanglement generation can not happen, we are never able to increase fidelities to levels they once were.}
\end{tcolorbox}  

Next, we wanted to show the fidelity degradation over increasing distances while keeping total node count constant. This allows us to observe scaling from another perspective, showing a more gradual decrease at lower distance which comes about due to the individual node separations.  

\begin{figure}[H]
    \centering
    \includegraphics[width=8cm]{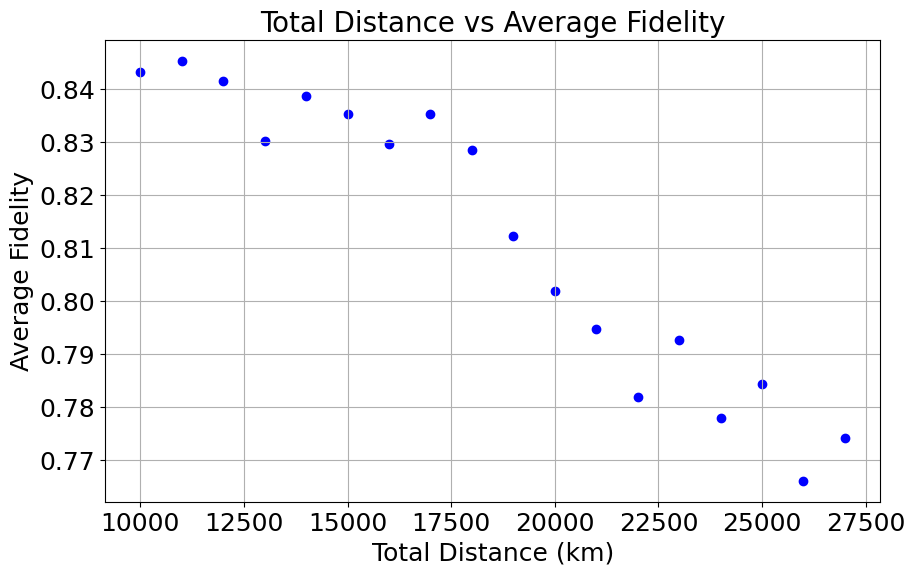}
    \caption{End-to-end fidelities for a fixed repeater count of 10, while scaling network distance. Non-linear downward trend suggesting effects from both decoherence and swapping failures.}
    \label{Figure 4}
\end{figure}

Fidelity remains stable for a while, approximately between 0.83 and 0.84, up to around 17,500 km, indicating that our network successfully maintains entanglement quality over moderate distances. Beyond this threshold, however, there is a much quicker decline in fidelity; it drops below 0.82 and dips to approximately 0.78 for distances exceeding 20,000 km.
This degradation in fidelity beyond 17,500 km suggests a practical limit of our simulated network. The accelerated decrease past 20,000 km results from intensified quantum decoherence and loss errors, underlining the need for alternate quantum communication strategies.

\begin{tcolorbox}[breakable,width=0.48\textwidth,title={Takeaway:},boxrule=.3mm,colback=white,coltitle=white,left=.5mm, right=.5mm, top=.5mm, bottom=.5mm]    
   \emph{In large networks of fixed node count, fidelity decreases non-linearly as the total distance increases. This means that keeping networks below certain dropoff thresholds (outside of the maximum separation threshold) might be useful.}
\end{tcolorbox}  

Expanding these experiments to lower distances could reveal other important trends, and potentially the inflection point where the trend shifts to experience more significant degradation effects.


These experiments operate in a simulated, optimized, and idealized environment that does not mirror practical implementations of quantum communication networks. Expanding our simulations to match the conditions of a real network could potentially be beneficial, but having identified strategies for scaling quantum repeater networks and for profiling their limitations, we will be able to collect data on physically implemented networks more systematically. 

\section{Discussion}

In a heterogeneous network, where some nodes are separated by large distances (where decoherence is the primary limiting factor) and others by just a few kilometers, the total number of nodes may not accurately reflect the network's performance. Surprisingly, a network with shorter inter-node distances could perform worse if fidelities are not sufficiently maintained or if swaps and generation attempts do not occur with high enough probability. These are issues that are often challenging to control with NISQ technologies. For instance, in a network of 10 nodes spread over 1000 km, adding more nodes might actually degrade performance, even if distance alone is not a problem.

One potential solution is to employ better routing algorithms or to plan the network more thoughtfully, with strategically planned routing and opportunistic requests\cite{9796816}. This approach could mitigate the challenges posed by decoherence and failure probabilities, enhancing the overall performance of the quantum network.

In odd-node count separation entanglement requests, the desynchronization problem could be solved by implementing quantum memories on the BSM nodes so that photons could be temporarily suspended until their counterparts arrive in order to buy time for synchronized joint measurements to be made. Aliro quantum \cite{aliro_quantum_memories_2024} notes that quantum memories are essential for practical quantum networks to operate. This is made apparent by our findings and is important to consider when implementing real-world quantum networks. 

Different types of qubits could be used within our networks to make them more robust to decoherence and easier to perform entanglement swapping and generation operations. One such type of qubit that is useful because of its extremely high coherence time and ability to exist at room temperature is nitrogen-vacancy (NV) centers in diamond. Developing networks based on NV centers could significantly increase scalability and reliability \cite{Ruf2021Quantum, Rozpedek2019NearTerm}. Another example, if found practical, is topological qubits which hold the promise of being extremely robust and resistant to noise. This type of qubit could revolutionize quantum computing and quantum communications by providing ideal quantum states for successful operations \cite{Aguado2020Majorana, Wang2022Entangled}.

Real-world implementation of network topologies beyond simple chains is highly anticipated. Several studies, including the application of game theory to optimize network structures, have highlighted the potential of diverse topologies \cite{Dey2020Quantum}. While principles of scalability in chain networks can be broadly applied, exploring optimal configurations in varying topologies remains crucial for future advancements.

\section{Conclusion}
This investigation into the scalability of quantum communication networks reveals significant insights into the limitations and potential of quantum network technologies. Our initial analysis of homogeneous quantum networks demonstrated a predictable linear scalability, leading to more sophisticated simulations of heterogeneous networks through the SeQUeNCe framework. These simulations, which account for the effects of a variety of network parameters including node separation and spacing, provided a nuanced understanding of the challenges in scaling quantum networks, particularly regarding entanglement generation and fidelity. The research highlights the delicate balance between the number of repeaters used in a network, individual node distance separation, and the overall network performance, identifying critical points that could guide the optimization of quantum communication networks. Future work will explore the effects of changing network topologies, and heterogeneous repeater architectures, and will expand network routing protocols based on our findings. Additionally, potential modifications in qubit types and implemented technologies will be considered to enhance network performance further. This paper contributes to the quantum computing field by delineating a path towards more scalable and efficient quantum communication networks, crucial for the next generation of quantum technology applications.

\bibliographystyle{IEEEtran}

\end{document}